\documentclass[a4paper]{article}
\usepackage{ISCSLP2026}
\usepackage{ifthen}
\usepackage{comment}
\usepackage{pifont}
\usepackage{multirow}
\usepackage{booktabs} 

\title{Phoenix-VAD: Streaming Semantic Endpoint Detection for Full-Duplex Speech Interaction}
\name{
    Weijie Wu$^1$, Wenhao Guan$^2$, Kadi Wang$^1$, Peijie Chen$^1$, \\
    Zhuanling Zha$^3$, Junbo Li$^3$, Jun Fang$^3$, Lin Li$^{2*}$, Qingyang Hong$^{1*}$
    \thanks{\quad $^{*}$ Corresponding author.}
}

\address{
    $^1$ School of Informatics, Xiamen University, China \\
    $^2$ School of Electronic Science and Engineering, Xiamen University, China \\
    $^3$ DiDi Global Inc., Beijing, China
}

\email{
    vjjjjjj@stu.xmu.edu.cn
}

\begin{document}

\maketitle
\begin{abstract}
   Spoken dialogue models have significantly advanced intelligent human-computer interaction. However, existing systems lack a modular full-duplex prediction mechanism for semantic endpoint detection, hindering seamless turn-taking. In this paper, we introduce Phoenix-VAD, a streaming semantic endpoint detection framework based on Large Language Models (LLMs). Specifically, Phoenix-VAD leverages the semantic comprehension capability of the LLM and a sliding window training strategy to achieve reliable semantic endpoint detection while supporting streaming inference. Extensive experimental results on both synthetic and real-world datasets demonstrate that Phoenix-VAD achieves competitive performance across semantically complete and incomplete scenarios. By bridging the gap between streaming inference and semantic comprehension, Phoenix-VAD provides reliable and flexible support for next-generation human–computer interaction.
\end{abstract}

\section{Introduction}

Spoken dialogue models represent one of the most direct methods of human-computer interaction. Recent advancements in spoken dialogue models, exemplified by systems like GPT-4o, have captured significant attention in the speech domain~\cite{qwen-omni, kimi-audio, vita-audio, step-audio, SALMONN, qian2023polyvoice}. In human–computer interaction, full-duplex communication is critical, as users can interrupt the voice assistant at any time. Traditional full-duplex dialogue systems relied on simple Voice Activity Detection (VAD) to determine whether the user intended to interrupt~\cite{cleans2s,x-talk, chen2025fireredchat, huang2024audiogpt}. Upon detecting user speech, the system controls the model response through a WebSocket interface. However, this approach only distinguishes between speech and silence~\cite{Formant-based-vad, energy-vad}, operating without the context of the user’s state and intention. Recent works address this by incorporating semantic understanding into full-duplex dialogue systems. Operating as a cascaded system, Semantic VAD~\cite{semantic-vad} leverages an auxiliary Automatic Speech Recognition (ASR) module and a lightweight LLM to determine whether a user’s utterance is semantically complete. Likewise, TEN Turn Detection~\cite{TEN_Turn_Detection}, designed as a text-only semantic judgment model, relies on an external ASR front-end to transcribe speech before performing analysis. To reduce model complexity, Smart Turn v3\footnote{https://huggingface.co/pipecat-ai/smart-turn-v3} utilizes Whisper Tiny~\cite{whisper} as a backbone with a linear classification layer. To achieve unified modeling, Moshi~\cite{kyutai2024moshi} integrates a base LLM with a smaller transformer to support real-time streaming predictions, forming a coupled architecture that requires retraining for deployment across different dialogue models. Concurrently, some works~\cite{RTTL-DG, lslm, wang2025ntpp, dgslm} fuse information from both channels for autoregressive generation and turn-taking detection. Other approaches~\cite{freeze-omni, chen2025minmo, mini-omni2} introduce an additional linear layer to predict whether the user intends to interrupt. Therefore, current methodologies typically face a structural trade-off between achieving strong semantic comprehension without cascaded ASR dependencies and maintaining a plug-and-play architecture that supports streaming inference.

\begin{figure}[!t]
  \centering
  \includegraphics[width=0.85\linewidth]{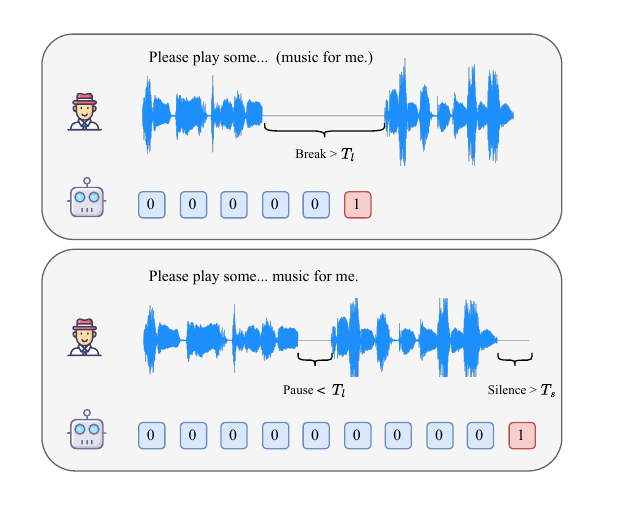}
  
  \caption{Illustration of semantic endpoint detection, where 0 and 1 represent \textit{Continue Speaking} and \textit{Stop Speaking}, respectively. $T_l$ and $T_s$ correspond to the timeout thresholds for semantically incomplete and complete queries.}
  \label{fig:wav_case}
\end{figure}


\begin{table*}[h]
\centering
\small
\caption{Feature comparison of different models}
\label{tab:model_comparison}
\begin{tabular}{lcccc}
\toprule
Model & ASR-Free & Plug-and-Play & Streaming Inference & Semantic Capability \\
\midrule
Acoustic VAD~\cite{Formant-based-vad, energy-vad,SileroVAD} & \ding{51} & \ding{51} & \ding{51} & \ding{55} \\
Cascaded Systems~\cite{semantic-vad, TEN_Turn_Detection} & \ding{55} & \ding{51} & \ding{55} & \ding{51} \\
Smart Turn v3 & \ding{51} & \ding{51} & \ding{55} & \ding{51} \\
Phoenix-VAD (Ours) & \ding{51} & \ding{51} & \ding{51} & \ding{51} \\
\bottomrule
\end{tabular}
\end{table*}

To provide a decoupled and pluggable full-duplex predictor without relying on external ASR front-ends, we propose Phoenix-VAD, a semantic endpoint detection model based on LLM. Phoenix-VAD features a plug-and-play design, enabling direct deployment across different dialogue models without additional training. As shown in Table \ref{tab:model_comparison}, by combining a sliding window strategy with the semantic comprehension capability of LLM, Phoenix-VAD achieves semantic understanding while supporting streaming inference. Moreover, Phoenix-VAD adopts a standard Speech Language Model (SLM), eliminating the need for an extra ASR module and enabling direct modeling of speech. Consequently, this direct audio processing mitigates the latency overhead and acoustic information loss inherently associated with cascaded architectures. As shown in Figure \ref{fig:wav_case}, Phoenix-VAD performs endpoint detection based on the semantic completeness of the user’s speech and determines termination using different timeout thresholds. When the user’s query is semantically incomplete, the model applies a longer timeout threshold to avoid premature termination of the response. To this end, our contributions can be summarized as follows:

\begin{itemize}
\item We propose Phoenix-VAD, a plug-and-play LLM-based model that performs streaming semantic endpoint detection for full-duplex speech interaction.
\item We adopt a sliding window training strategy, allowing the model to capture semantic information while supporting streaming inference.
\item Experimental results demonstrate that Phoenix-VAD achieves competitive performance in both semantically complete and semantically incomplete scenarios.

\end{itemize}

\section{Method}
\label{sec:method}

\subsection{Overall structure}
\label{ssec:Overview}

\begin{figure}[!t]
  \centering
  \includegraphics[width=0.78\linewidth]{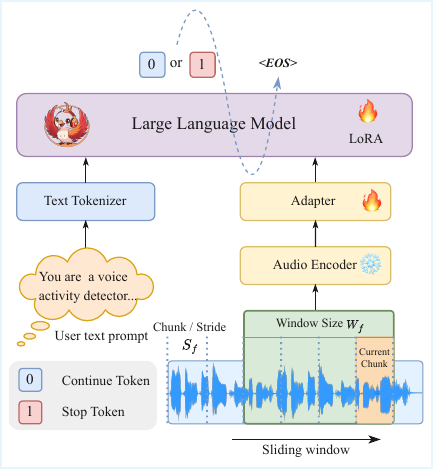}
  \caption{The architecture of the Phoenix-VAD model.}
  \label{fig:Phoenix-VAD}
\end{figure}

Phoenix-VAD is an autoregressive LLM-based architecture for semantic endpoint detection. An overview of the architecture is provided in Figure~\ref{fig:Phoenix-VAD}. Specifically, Phoenix-VAD consists of three core components:

\begin{itemize}
    \item \textbf{Audio Encoder:} We employ a Zipformer encoder~\cite{zipformer} with approximately 150M parameters to process raw waveform inputs and generate frame-level audio features $\mathbf{A} = [a_1, ..., a_N]$ at 25 Hz. The encoder is pretrained on over 100k internal ASR data for robust representations.
    
    \item \textbf{Modality Adapter:} To bridge the modality gap between audio and text, we utilize an adapter composed of two linear layers with a ReLU activation function. The adapter transforms feature dimensions and downsamples temporally to align audio features with the text embedding space. Inspired by \cite{slam-asr,llama-omni,slam-omni}, we downsample $\mathbf{A}$ by concatenating every $k$ consecutive frames along the feature dimension, yielding intermediate features $\mathbf{A}^I = [a_1^I, a_2^I, \ldots, a_{N'}^I]$, where $a_i^I = a_{(i-1) \cdot k + 1} \oplus a_{(i-1) \cdot k + 2} \oplus \cdots \oplus a_{i \cdot k}$ and $N' = \left\lfloor \frac{N}{k} \right\rfloor$. The intermediate features $\mathbf{A}^I$ are then projected through the adapter, resulting in the adapted representation $\mathbf{A}^P = W_2 \bigl( \mathit{ReLU}(W_1 \mathbf{A}^I + b_1) \bigr) + b_2$, where $W_1, W_2$ are weight matrices and $b_1, b_2$ are bias vectors.
    
    \item \textbf{LLM Backbone:} Finally, we use Qwen2.5-0.5B-Instruct\footnote{https://huggingface.co/Qwen/Qwen2.5-0.5B-Instruct} as the backbone, which takes the adapted features \(\mathbf{A}^P\) concatenated with text prompt embeddings \(\mathbf{T}^P\) as input to predict the semantic completeness state of the input speech segment.
\end{itemize}

\subsection{Sliding window training strategy}
\label{ssec:sliding_window}
To enable streaming inference, we employ a sliding window training strategy that makes predictions using only the audio within each window, reducing dependence on the entire input sequence. Compared with processing the entire sequence, it enables incremental chunk-wise predictions, offering potential advantages in latency. Meanwhile, the model can leverage the information within each window, providing sufficient local context for semantic reasoning while seamlessly supporting streaming inference.

During training, we apply the sliding window strategy to the Fbank sequence $\mathbf{F} = [f_1, f_2, \ldots, f_N]$, with a frame rate of 100 Hz. For chunk index $c$, the window spans frames from $\ell_c = \max\{1, \min(c \cdot S_f, N) - W_f\}$ to $r_c = \min(c \cdot S_f, N)$, defining the window $\mathbf{F}_c = [f_{\ell_c}, f_{\ell_c+1}, \ldots, f_{r_c}]$, 
where $S_f$ and $W_f$ denote the chunk stride and window size in frames. 

Following Section~\ref{ssec:Overview}, Zipformer encoder converts the sequence $\mathbf{F}_c$ to 25Hz, resulting in the encoder output $\mathbf{A}_c$. The adapter then maps $\mathbf{A}_c$ to intermediate features $\mathbf{A}_c^I$ and the adapted representation $\mathbf{A}_c^P$, which are subsequently provided as input to LLM.




\subsection{Training objective}
\label{ssec:training_objective}
For each window sequence, supervision is applied only to the last chunk, with the target
\begin{equation}
\label{eq:label_seq}
Y_c = \{y_c, \langle\mathrm{eos}\rangle\}, \quad y_c \in \{0,1\},
\end{equation}
where $y_c=0$ and $y_c=1$ denote \textit{Continue Speaking} and \textit{Stop Speaking}, respectively. The $\langle\mathrm{eos}\rangle$ token marks the end of autoregressive decoding.

We adopt VocalNet~\cite{vocalnet} as the base framework. During training, the audio encoder is frozen, while the adapter and LLM are jointly optimized, with LoRA applied to the LLM. The model is trained using the standard cross-entropy loss:
\begin{equation}
\mathcal{L}(\theta) = -\sum_{t=1}^{|Y_c|} \log p_\theta \big(Y_c^{(t)} \mid Y_c^{(<t)}, \mathbf{A}_c^P, \mathbf{T}^P \big).
\end{equation}

\section{Data Preparation}
\label{ssec:Data preparation}

In conversational interactions, overlapping speech and interruptions are common. For human-computer interaction systems, when interruptions occur, we prioritize user experience in our design by allowing only user-initiated interruptions. Semantic endpoint detection in this context is formulated as a user state detection task, where the model determines whether the user has finished speaking using both acoustic and semantic information. Therefore, we define two labels as follows:

    
    


\begin{itemize}
    \item \textbf{Continue Speaking} (\texttt{<|C-S|>}): If the user's query is incomplete, the model continues processing audio until the user stops speaking or is interrupted.
    
    \item \textbf{Stop Speaking} (\texttt{<|S-S|>}): The model stops recording and processes the utterance with timeout thresholds based on semantic completeness. A short timeout $T_s$ is applied to semantically complete queries to reduce response latency, whereas a longer timeout $T_l$ is applied to incomplete queries to prevent premature interruption.
\end{itemize}

Data preparation consists of three stages: text generation, speech synthesis, and timestamp annotation. To begin with, semantically complete and incomplete text samples are generated by combining internal textual resources with the ChatGPT API. Then, speech samples are synthesized using Index-TTS~\cite{indextts}. To enhance speaker diversity, vocal timbre is controlled by randomly sampling speaker prompts from a timbre library, which contains 1,007 English and 1,010 Chinese human audio prompts sourced from seed-tts-eval~\cite{seed-tts}. Hesitations and interruptions are simulated by inserting silent segments at specified text locations. Finally, we use Paraformer~\cite{paraformer} to annotate timestamps for each character and generate the final training labels by extracting the \textit{stop speaking} timestamps.

\section{Experiments}
\label{sec:Experiments}
\subsection{Dataset and settings}

We use the dataset constructed in Section~\ref{ssec:Data preparation} for training, which contains approximately 400,000 audio samples totaling around 570 hours. Additionally, to comprehensively evaluate the semantic endpoint detection capability, we established two distinct test sets: a synthetic set and a real-world set. For the synthetic test set, we follow the generation procedure in Section~\ref{ssec:Data preparation} to synthesize a collection comprising 2,000 samples for each semantic category. To further assess robustness in realistic scenarios, we curated a real-world test set by recruiting five volunteers to record queries in environments with background noise, yielding approximately 110 samples per category to simulate user behavior in authentic acoustic conditions. All annotations passed consistency verification to ensure label reliability. Regarding implementation, the model was fine-tuned for one epoch on 32 NVIDIA A100 80 GB GPUs with a batch size of 64 and a learning rate of $5\times 10^{-5}$, using a cosine annealing schedule with 0.03 warmup ratio. During training, we employ a sliding window strategy where we set $W_f = 256$ and $S_f = 32$, corresponding to a window size of 2560 ms and a stride of 320 ms, respectively. The timeout thresholds are set to $T_s = 400\,\text{ms}$ and $T_l = 1000\,\text{ms}$, based on empirical observations of typical pause lengths in speech interactions.

\begin{table}[!t]
  \centering
  \small
  \renewcommand{\arraystretch}{1.12} 
  \setlength{\tabcolsep}{4pt}
  
  \caption{Chunk-level confusion matrices on synthetic datasets for semantically complete and incomplete queries. \textit{GT} and \textit{Est} denote Ground Truth and Estimation, respectively.}
  \label{t2}
  
  \resizebox{\columnwidth}{!}{
  \begin{tabular}{l|cc|cc}
    \toprule
    \multirow{2}{*}{\textit{GT} \textbackslash \textit{Est}} & \multicolumn{2}{c|}{\textbf{Complete Set}} & \multicolumn{2}{c}{\textbf{Incomplete Set}} \\
    \cmidrule(lr){2-3} \cmidrule(lr){4-5}
     & \texttt{<|S-S|>} & \texttt{<|C-S|>} & \texttt{<|S-S|>} & \texttt{<|C-S|>} \\
    \midrule
    \texttt{<|S-S|>}     & 1775 & 225 & 1784 & 216 \\
    \texttt{<|C-S|>} & 148 & 24826 & 101 & 19662 \\
    \midrule
    \textbf{Recall}    & \multicolumn{2}{c|}{0.888} & \multicolumn{2}{c}{0.892} \\
    \textbf{Precision} & \multicolumn{2}{c|}{0.923} & \multicolumn{2}{c}{0.946} \\
    \textbf{F1 Score}  & \multicolumn{2}{c|}{0.905} & \multicolumn{2}{c}{0.918} \\
    \textbf{Accuracy}  & \multicolumn{2}{c|}{0.986} & \multicolumn{2}{c}{0.985} \\
    \bottomrule
  \end{tabular}
  }
\end{table}

\subsection{Evaluation results}


\subsubsection{Chunk-level evaluation}


We first evaluated the performance of Phoenix-VAD on the synthetic test set. 
Following the same processing pipeline as in the training phase, each audio sample was segmented into fixed-length chunks, with the model predicting a state token for each segment. We constructed confusion matrices and computed standard performance metrics, including Recall, Precision, F1 score, and Accuracy. As shown in Table \ref{t2}, the evaluation results validate the effectiveness of the model in both semantically incomplete and semantically complete scenarios. Specifically, the model exhibits robust performance across these conditions, demonstrating particularly high reliability in identifying \textit{Continue Speaking} segments. Although a slight performance variance is observed in \textit{Stop Speaking} scenarios, which is a common challenge attributed to the inherent variability of semantic endpoint detection in spontaneous speech, the model maintains a high level of overall performance.

\begin{table*}[!t]
  \centering
  \small
  \caption{Performance comparison on Synthetic and Real-world datasets. The evaluation metric is \textbf{sentence-level accuracy}. For Phoenix-VAD, we apply a \textbf{strict correctness criterion}: a sentence is correct \textbf{if and only if} all chunk-level predictions match the ground truth labels. ``Stream'' denotes support for streaming inference. Bold indicates the best performance.}
  \label{t4}
  \setlength{\tabcolsep}{9pt} 
  \begin{tabular}{l c c c c c c}
    \toprule
    \multirow{2}{*}{\textbf{Model}} & \multirow{2}{*}{\textbf{Params}} & \multirow{2}{*}{\textbf{Stream}} & \multicolumn{2}{c}{\textbf{Synthetic Set}} & \multicolumn{2}{c}{\textbf{Real-world Set}} \\
    \cmidrule(lr){4-5} \cmidrule(lr){6-7}
    & & & \textbf{Complete} & \textbf{Incomplete} & \textbf{Complete} & \textbf{Incomplete} \\
    \midrule
    TEN Turn Detection & 7.6B & \ding{55} & \textbf{0.929} & 0.859 & \textbf{0.930} & 0.829 \\
    Paraformer + TEN Turn Detection & 7.8B & \ding{55} & 0.919 & 0.794 & 0.887 & 0.829 \\
    Smart Turn v3 & 8M & \ding{55} & 0.800 & 0.561 & 0.704 & 0.623 \\
    \midrule
    \textbf{Phoenix-VAD (Ours)} & 0.6B & \ding{51} & 0.888 & \textbf{0.892} & 0.800 & \textbf{0.838} \\
    \bottomrule
  \end{tabular}
\end{table*}

\begin{table}[!t]
  \centering
  \small
  \caption{Results on the Easy Turn real-world test set.}
  \label{easyturn_test_set}
  \setlength{\tabcolsep}{4pt}
  \begin{tabular}{l c c}
    \toprule
    \textbf{Model} & \textbf{Complete} & \textbf{Incomplete} \\
    \midrule
    Easy Turn & 0.947 & 0.973 \\
    Smart Turn v3 & 0.747 & 0.360 \\
    Paraformer + TEN Turn Detection & 0.840 & 0.893 \\
    \midrule
    \textbf{Phoenix-VAD (Ours)} & 0.847 & 0.867 \\
    \bottomrule
  \end{tabular}
\end{table}

\begin{table}[!t]
  \centering
  \small
  \caption{Ablation study of Phoenix-VAD.}
  \label{t5}
  \setlength{\tabcolsep}{4pt}
  \begin{tabular}{l c c}
    \toprule
    \textbf{Method} & \textbf{Complete} & \textbf{Incomplete} \\
    \midrule
    \textbf{Phoenix-VAD (Proposed)} & \textbf{0.905} & \textbf{0.918} \\
    \midrule
    \quad w/ smaller chunk size (160 ms) & 0.798 & 0.839 \\
    \quad w/o joint training & 0.208 & 0.494 \\
    \bottomrule
  \end{tabular}
\end{table}

\subsubsection{Sentence-level evaluation}

To provide a more comprehensive assessment of Phoenix-VAD in realistic interaction scenarios, we perform a systematic sentence-level evaluation on both synthetic and real-world datasets. As detailed in Table~\ref{t4}, we compare Phoenix-VAD against several recent baseline systems. Specifically, TEN Turn Detection is evaluated under two distinct configurations: first, utilizing ground-truth text, and second, employing a cascaded pipeline that integrates Paraformer as an ASR front-end. It is important to note that the majority of existing open-source baselines are limited to non-streaming inference, necessitating access to the full speech context to formulate a one-time semantic decision. To ensure a fair comparison, we aggregate the chunk-level predictions of Phoenix-VAD into sentence-level decisions, enforcing a strict correctness criterion: a sentence is considered correct \textbf{if and only if} all chunk-level predictions match the ground truth labels.

Despite this rigorous protocol, Phoenix-VAD demonstrates competitive performance against substantially larger baselines. In particular, it consistently outperforms all baselines on semantically incomplete sentences across both datasets, reducing false interruptions and enabling smoother turn-taking. On semantically complete sentences, it maintains accuracy comparable to non-streaming approaches. To further evaluate generalization, we benchmark Phoenix-VAD on the out-of-domain Easy Turn real-world test set~\cite{easyturn}, as reported in Table~\ref{easyturn_test_set}. Although trained exclusively on synthetic data and being the only streaming model among the compared systems, Phoenix-VAD still substantially outperforms Smart Turn v3 and remains comparable to the 7.8B cascaded system.

\subsection{Latency analysis}



Compared to ASR-based cascaded systems, Phoenix-VAD benefits from streaming inference, demonstrating a significant advantage in latency. We denote the total duration of the user's speech by $T_{\text{speech}}$ and the model processing overhead by $t$. Phoenix-VAD operates with a sliding window stride of $T_{\text{stride}} = 320\,\text{ms}$, satisfying $T_{\text{stride}} \ll T_{\text{speech}}$. Notably, per-window compute is fixed and does not grow with utterance length. Typically, ASR-based cascaded systems cannot initiate the processing pipeline until the full speech input is received, resulting in a first-packet latency of $T_{\text{speech}} + t_{\text{cascade}}$. In contrast, Phoenix-VAD performs inference and generates a VAD decision immediately after receiving an initial audio segment of duration $T_{\text{stride}}$, reducing the first-packet latency to merely $T_{\text{stride}} + t_{\text{Phoenix-VAD}}$.


We evaluated the model processing times on a single NVIDIA A6000 48 GB GPU. 
The cascaded system, composed of Paraformer and TEN Turn Detection, exhibited a processing time of $t_{\text{cascade}} = 284\,\text{ms}$, while Phoenix-VAD required only $t_{\text{Phoenix-VAD}} = 140\,\text{ms}$, indicating a reduction of approximately 51\% in processing latency. 
Assuming a standard human-computer interaction scenario with a user query duration $T_{\text{speech}}$ of 3 seconds, the overall latency of the cascaded system reaches $3000 + 284 = 3284\,\text{ms}$. 
Conversely, the latency of Phoenix-VAD is determined by the stride length rather than the full utterance, totaling $320 + 140 = 460\,\text{ms}$. 
This represents a substantial reduction of approximately 86\% in overall latency.



\subsection{Ablation studies}
\label{ssec:ablation}
We conduct an ablation study on the synthetic set to analyze the impact of critical design choices, as shown in Table~\ref{t5}. We first reduce the chunk size from 320 ms to 160 ms to evaluate the influence of finer temporal granularity on model performance. In addition, to validate the necessity of joint training, we employ a two-stage modular training strategy: the adapter is pretrained on internal ASR data and then frozen, and only the LoRA parameters are optimized in the second stage. Results show that reducing the chunk size negatively impacts performance, as finer temporal granularity increases prediction uncertainty near decision boundaries and magnifies the effect of timestamp errors produced by Paraformer. Moreover, training the adapter solely on ASR data yields inferior performance compared to our proposed approach, as this supervision emphasizes lexical alignment while providing limited exposure to the temporal cues required for accurate speech boundary detection, thereby leaving the adapter less sensitive to fine-grained timing.

\section{Conclusions and Future Work}
\label{sec:Conclusions and future work}


In this paper, we introduce Phoenix-VAD, an LLM-based model for semantic endpoint detection. By leveraging the semantic comprehension capability of LLM and a sliding window training strategy, Phoenix-VAD enables reliable user speech state detection while supporting streaming inference. Experiments show that Phoenix-VAD achieves reliable semantic endpoint detection, enhancing full-duplex detection in speech interactions. In future work, we plan to develop the rejection capability of Phoenix-VAD, scale up training with large-scale real-world recordings, extend its context awareness to support multi-turn dialogue understanding by modeling cross-turn semantic dependencies, and apply it to end-to-end dialogue systems.


\section{Acknowledgements}
This work was supported in part by the National Natural Science Foundation of China under Grants 62276220 and 62371407 and the Innovation of Policing Science and Technology, Fujian Province (Grant number: 2024Y0068)

\bibliographystyle{IEEEtran}

\bibliography{mybib}


\end{document}